\documentclass[aps,prl,twocolumn,showpacs,byrevtex]{revtex4}

\usepackage{times,mathptm}
\usepackage[dvips]{graphicx,color}
\usepackage{titlesec}
\newcommand{\bqa}{\begin{eqnarray*}}
\newcommand{\eqa}{\end{eqnarray*}}
\makeatletter
\renewcommand{\@seccntformat}[1]{}
\makeatother

\begin{document}
\title{Origin of enhanced chemical precompression in cerium hydride CeH$_9$}
\author{Hyunsoo Jeon, Chongze Wang, Seho Yi, and Jun-Hyung Cho$^{*}$}
\affiliation{Department of Physics, Research Institute for Natural Science, and HYU-HPSTAR-CIS High Pressure Research Center, Hanyang
University, 222 Wangsimni-ro, Seongdong-Ku, Seoul 04763, Republic of Korea}
\email{chojh@hanyang.ac.kr}

\begin{abstract}
The rare-earth metal hydrides with clathrate structures have been highly attractive because of their promising high-$T_{\rm c}$ superconductivity at high pressure. Recently, cerium hydride CeH$_9$ composed of Ce-encapsulated clathrate H cages was synthesized at much lower pressures of 80$-$100 GPa, compared to other experimentally synthesized rare-earth hydrides such as LaH$_{10}$ and YH$_6$. Based on density-functional theory calculations, we find that the Ce 5$p$ semicore and 4$f$/5$d$ valence states strongly hybridize with the H 1$s$ state, while a transfer of electrons occurs from Ce to H atoms. Further, we reveal that the delocalized nature of Ce 4$f$ electrons plays an important role in the chemical precompression of clathrate H cages. Our findings not only suggest that the bonding nature between the Ce atoms and H cages is characterized as a mixture of ionic and covalent, but also have important implications for understanding the origin of enhanced chemical precompression that results in the lower pressures required for the synthesis of CeH$_9$.
\end{abstract}

\maketitle

\section*{Introduction}
In recent years, rare-earth metal hydrides have attracted much attention due to the possibility of their realization of room-temperature superconductivity (SC)~\cite{review-Zurek,review-Eremets,ExpLaH10-PRL2019,ExpLaH10-Nature2019,ExpCeH9-Nat.Commun2019T.Cui, ExpCeH9-Nat.Commun2019-J.F.Lin,ExpYH6-arXiv2019-Eremet,ExpYH6-arXiv2019-Oganov}. First-principles density-functional theory (DFT) calculations together with the Migdal-Eliashberg formalism have predicted that the rare-earth metal hydrides such as yttrium hydride YH$_{10}$ and lanthanum hydride LaH$_{10}$ host a room-temperature SC at megabar pressures~\cite{rare-earth-hydride-PRL2017, rare-earth-hydride-PANS2017}, the origin of which is based on phonon-mediated electron pairing~\cite{BCS}. Subsequently, such a conventional SC of LaH$_{10}$ was experimentally observed with a superconducting transition temperature $T_{\rm c}$ of 250$-$260 K at a pressure of ${\sim}$170 GPa~\cite{ExpLaH10-PRL2019, ExpLaH10-Nature2019}. This $T_{\rm c}$ record of LaH$_{10}$ has been the highest temperature so far among experimentally available superconducting materials including cuprates~\cite{Cuprate-Nature1993, Cuprate-Rev-Nature2010} and Fe-based superconductors~\cite{Fe-based-Zhao2008, Fe-based-Chen2009}. Therefore, the experimental realization of room-temperature SC in LaH$_{10}$ has stimulated interests of the high-$T_{\rm c}$ community towards compressed metal hydrides under high pressure~\cite{PrH9-T.Cui-2020,NdHx-JACS2020,DyHx-Inorg.Chem.2020,H3S-phase-diagram-NC2019,Tech-Science2019,Li2MgH16-PRL2019,Li6P-PRL2019}.

However, since the synthesis of LaH$_{10}$ was performed at ${\sim}$170 GPa~\cite{ExpLaH10-PRL2019,ExpLaH10-Nature2019}, it has been quite demanding to discover H-rich rare-earth hydrides synthesized at a moderate pressure below ${\sim}$100 GPa, which is easily and routinely achievable in the diamond anvil cell (DAC)~\cite{diamondanvil-Rev2009-Bassett,diamondanvil-Rev2018-K.K.Mao}. Recently, two experimental groups~\cite{ExpCeH9-Nat.Commun2019T.Cui,ExpCeH9-Nat.Commun2019-J.F.Lin} reported the successful synthesis of cerium hydride CeH$_9$ at 80$-$100 GPa. Using X-ray diffraction measurements and DFT calculations~\cite{ExpCeH9-Nat.Commun2019T.Cui,ExpCeH9-Nat.Commun2019-J.F.Lin}, the crystal structure of CeH$_9$ was determined to be a hexagonal clathrate structure with the space group P6$_3$/$mmc$, where each Ce atom is surrounded by the H$_{29}$ cage consisting of 29 H atoms [see Fig. 1(a)]. It is remarkable that the H$-$H bond lengths in CeH$_9$ are close to those of solid metallic hydrogen that can be produced at high pressure above 400 GPa~\cite{MetalicH-Rev.Mod.Phys2012,MetalicH-PRL2015,MetalicH-PRB2016,MetalicH-Science2017}. Therefore, the discovery of CeH$_9$ having clathrate hydrogen networks suggests that the metallic state of solid hydrogen can be attained at relatively lower pressures by using binary hydrides with $f$-electron metals. It is noteworthy that the sizable reduction of H$-$H bond lengths in CeH$_9$ reflects the presence of a larger chemical precompression~\cite{chemically-precompressed-PRL2004,chemically-precompressed-PRL2006,chemically-precompressed-Science2008} compared to other rare-earth metal hydrides such as LaH$_{10}$ and YH$_{10}$~\cite{rare-earth-hydride-PRL2017,rare-earth-hydride-PANS2017,LaH10-PRB2019-Liangliang, LaH10-PRB2019-Chongze,LaH10-PRB2020-Chongze,YH10-Boeri,LaH10-Quamtum-crystal-Nature-2020}. However, the underlying mechanism of how the pressure required for the synthesis of CeH$_9$ is much reduced is yet to be identified.

\begin{figure}[ht]
\centering{ \includegraphics[width=8.0cm]{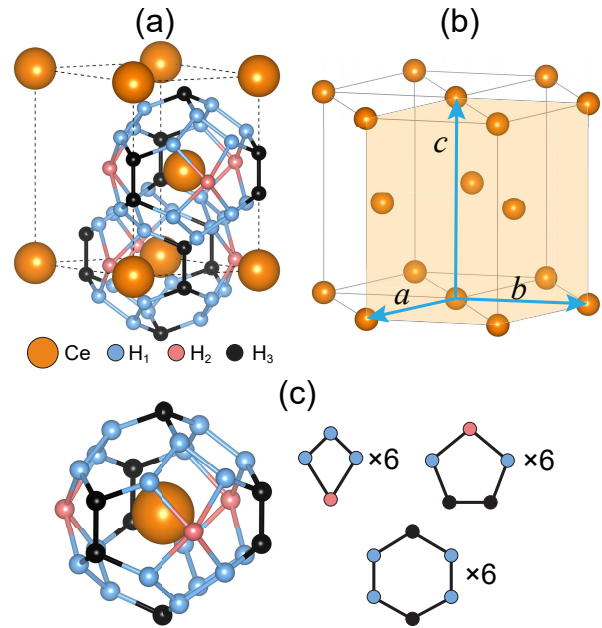} }
\caption{(a) Optimized structure of CeH$_{9}$ at 100 GPa and (b) hexagonal-close-packed (hcp) lattice of Ce atoms. Three different species of H atoms, H$_{1}$, H$_{2}$, and H$_{3}$, exist in H$_{29}$ cage. The isolated H$_{29}$ cage surrounding a Ce atom is displayed in (c), together with its constituent parts, i.e., six tetragon rings, six pentagon rings, and six hexagon rings.}
\end{figure}

In this paper, we investigate the electronic structure and bonding properties of CeH$_9$ at high pressure using first-principles DFT calculations with the inclusion of Hubbard on-site Coulomb interaction. The calculated band structure of CeH$_9$ shows a strong hybridization of the Ce 5$p$ semicore and 4$f$/5$d$ valence states with the H 1$s$ state. We reveal that the delocalized nature of Ce 4$f$ electrons contributes to yield a much larger chemical precompression of clathrate H$_{29}$ cages along the $c$ axis than in the $a$-$b$ plane. Despite a strong hybridization between the Ce- and H-derived electronic states, our Bader charge analysis shows a charge transfer from Ce to H atoms, thereby suggesting that the bonding nature between the Ce atoms and H$_{29}$ cages features the mixed ionic and covalent bonding. The present results provide new insight into understanding the underlying mechanism of the chemical precompression that requires relatively lower pressures for the synthesis of CeH$_9$~\cite{ExpCeH9-Nat.Commun2019T.Cui,ExpCeH9-Nat.Commun2019-J.F.Lin}, compared to other experimentally synthesized rare-earth hydrides LaH$_{10}$~\cite{ExpLaH10-PRL2019,ExpLaH10-Nature2019} and YH$_{6}$~\cite{ExpYH6-arXiv2019-Eremet,ExpYH6-arXiv2019-Oganov}.

\section*{Results and discussion}
\begin{figure}[ht]
\centering{ \includegraphics[width=8.0cm]{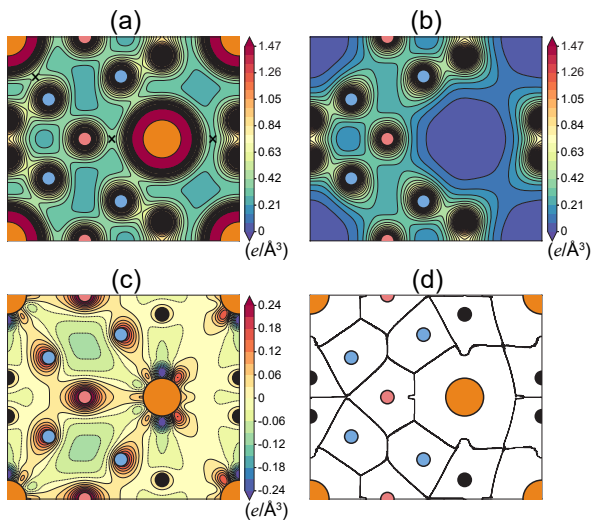} }
\caption{Calculated total charge density ${\rho}_{\rm tot}$ of (a) CeH$_9$ and (b) isolated H$_{29}$ cages. The saddle points of charge densities between Ce and H$_1$/H$_2$/H$_3$ atoms are marked ``${\times}$" in (a). The charge densities in (a) and (b) are plotted on the (110) plane with the contour spacings of 0.07 $e$/{\AA}$^3$. The charge density difference ${\Delta}{\rho}$ (defined in the text) is displayed in (c), with the contour spacing of ${\pm}$0.03 $e$/{\AA}$^3$. The Bader basins of Ce and H atoms are displayed in (d).}
\end{figure}

\begin{figure*}[h!t]
\centering{ \includegraphics[width=15.0cm]{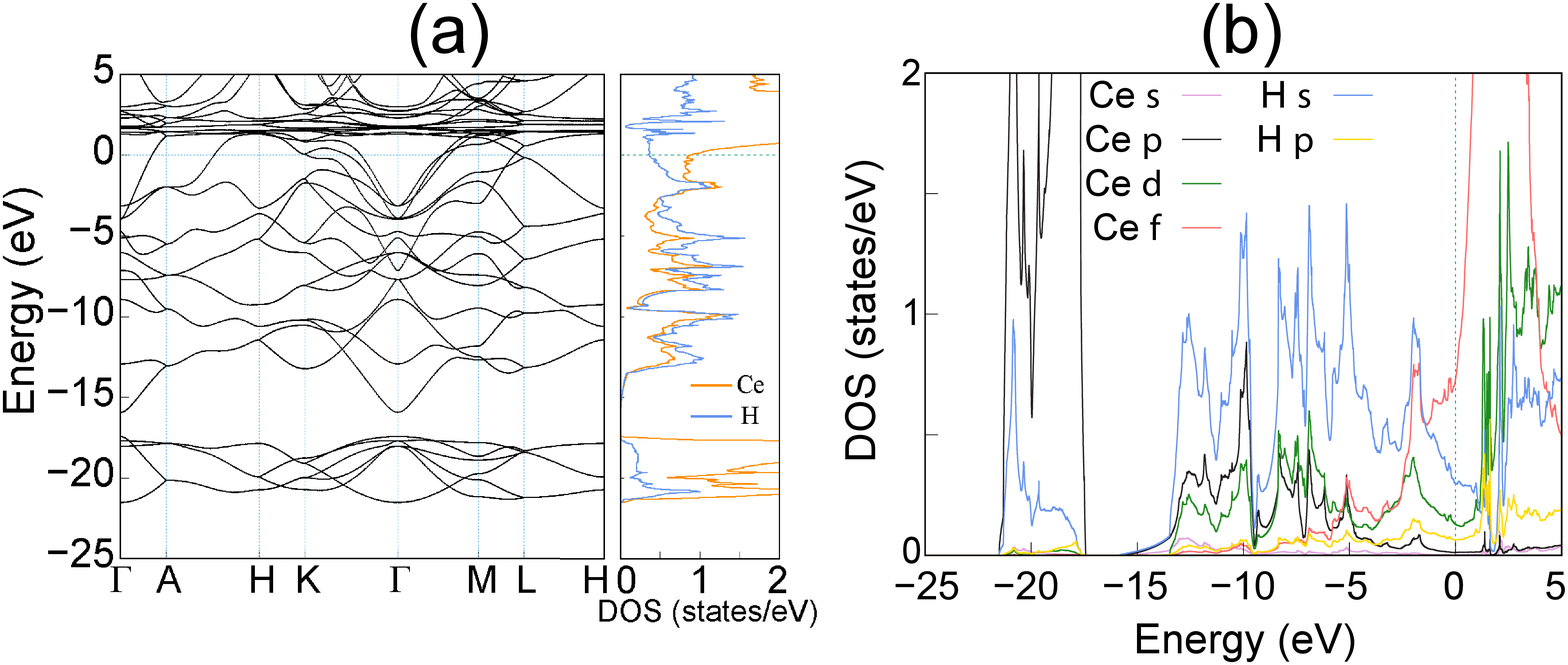} }
\caption{(a) Calculated band structure of CeH$_9$ together with the LDOS for Ce and H atoms. The energy zero represents $E_{\rm F}$. The PDOS of CeH$_9$ is also given in (b).}
\end{figure*}

We begin by optimizing the structure of CeH$_9$ using the PBE + U calculation. Figure 1(a) shows the optimized structure of CeH$_9$ at a pressure of 100 GPa, which is the same pressure employed in previous DAC experiments~\cite{ExpCeH9-Nat.Commun2019T.Cui,ExpCeH9-Nat.Commun2019-J.F.Lin}. Here, Ce atoms form a hcp lattice [see Fig. 1(b)] with the lattice constants $a$ = $b$ = 3.717 {\AA} and $c$ = 5.666 {\AA}, in good agreement with the experimental~\cite{ExpCeH9-Nat.Commun2019T.Cui,ExpCeH9-Nat.Commun2019-J.F.Lin} data of $a$ = $b$ = 3.66 {\AA} and $c$ = 5.58 {\AA}. Meanwhile, the H$_{29}$ cage surrounding a Ce atom is constituted by six tetragon rings, six pentagon rings, and six hexagon rings [see Fig. 1(c)]. Note that there are three species of H atoms [termed H$_1$, H$_2$, and H$_3$ in Fig. 1(a)] composing the H$_{29}$ cage. We find that the H$_{1}-$H$_1$, H$_{1}-$H$_2$, H$_{1}-$H$_3$, and H$_{3}-$H$_3$ bond lengths are 1.190, 1.486, 1.275, and 1.065 {\AA}, respectively. These H$-$H bond lengths in CeH$_9$ are close to those (0.98 and 1.21 {\AA}) predicted from metallic hydrogen at ${\sim}$500 GPa~\cite{MetalicH-PRB2016}. It is thus likely that the synthesized~\cite{ExpCeH9-Nat.Commun2019T.Cui,ExpCeH9-Nat.Commun2019-J.F.Lin} binary compound CeH$_9$ with the clathrate H$_{29}$ cages is able to generate H networks comparable to metallic hydrogen even at a low pressure of 100 GPa.

Figure 2(a) shows the calculated total charge density ${\rho}_{\rm tot}$ of CeH$_9$. It is seen that H atoms in the H$_{29}$ cage are bonded to each other with covalent-like bonding. Here, each H$-$H bond has a saddle point of charge density at its midpoint, similar to the C$-$C covalent bond in diamond~\cite{diamond}. The charge densities at the midpoints of the H$_{1}-$H$_2$, H$_{1}-$H$_3$, and H$_{3}-$H$_3$ bonds are 0.39, 0.56, and 0.85 $e$/{\AA}$^3$, respectively. These values in CeH$_9$ are larger than the corresponding ones [0.32, 0.45, and 0.76 $e$/{\AA}$^3$ in Fig. 2(b)] obtained from the isolated H$_{29}$ cages whose structure is taken from the optimized structure of CeH$_9$. This result implies that the H$-$H covalent bonds in CeH$_9$ are strengthened by a charge transfer from Ce to H atoms. Interestingly, the electrical charges of Ce and H atoms are connected to each other, reflecting a covalent-like bonding character. It is noteworthy that the charge densities at the points marked ``${\times}$" [in Fig. 2(a)] between Ce and H$_1$/H$_2$/H$_3$ atoms are close to that at the midpoint of the H$_{1}-$H$_2$ bond. This covalent nature of Ce$-$H bonds is attributed to a strong hybridization between the Ce and H electronic states, as discussed below.

\begin{figure*}[htb]
\centering{ \includegraphics[width=15.0cm]{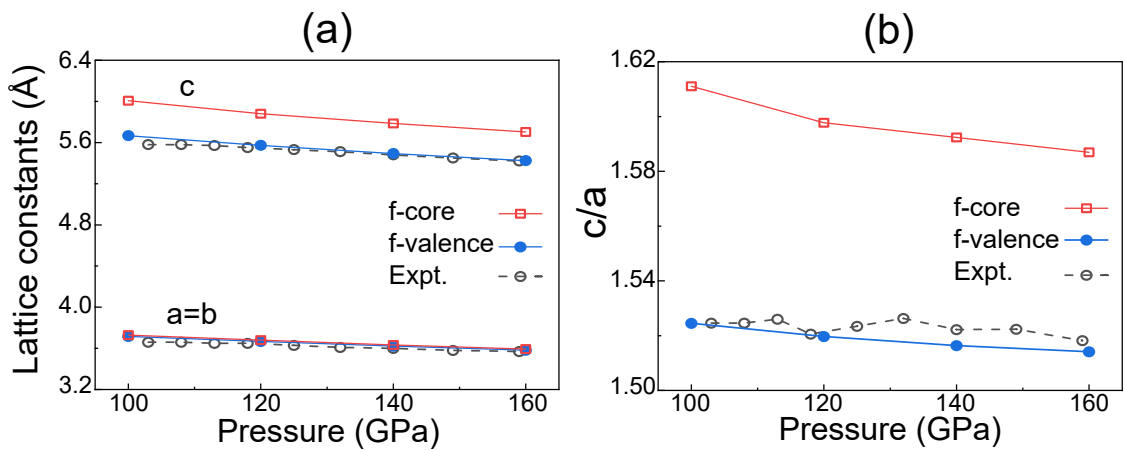} }
\caption{(a) Calculated lattice constants of CeH$_9$ as a function of pressure using the $f$-valence scheme, in comparison with those obtained using the $f$-core scheme and experiment~\cite{ExpCeH9-Nat.Commun2019T.Cui}. The resulting ${c/a}$ ratios as a function of pressure are also given in (b).}
\end{figure*}

To examine the charge transfer between Ce to H atoms, we calculate the charge density difference, defined as ${\Delta}{\rho}$ = ${\rho}_{\rm tot}$ $-$ ${\rho}_{\rm Ce}$ $-$ ${\rho}_{\rm H}$, where ${\rho}_{\rm Ce}$ and ${\rho}_{\rm H}$ represent the charge densities of the isolated Ce lattice [Fig. 1(b)] and the isolated H$_{29}$ cages [Fig. 2(b)], respectively. As shown in Fig. 2(c), ${\Delta}{\rho}$ illustrates how electronic charge is transferred from Ce to H atoms. It is seen that the charge transfer occurs mostly from Ce to H$_1$ and H$_2$. Meanwhile, the charge transfer from Ce to H$_3$ is minor. We further calculate the Bader charges~\cite{Bader} of CeH$_9$ to estimate the number of transferred electrons between Ce and H atoms. Figure 2(d) shows the Bader basins of the constituent atoms, obtained from the gradient of ${\rho}_{\rm tot}$~\cite{Bader}. We find that the calculated Bader charges of Ce, H$_1$, H$_2$, and H$_3$ basins are $-$9.47$e$ (including the 5$s^2$5$p^6$ semicore electrons), $-$1.34$e$, $-$1.31$e$, and $-$1.09$e$, respectively. Thus, we can say that Ce atoms lose electrons of 2.53$e$ per atom, while H$_1$, H$_2$, and H$_3$ atoms gain electrons of 0.34$e$, 0.31$e$, 0.09$e$ per atom, respectively.

In Figure 3(a), we display the calculated band structure of CeH$_9$, together with the local density of states (LDOS) for Ce and H atoms. The narrow bands located at ${\sim}$2 eV above $E_{\rm F}$ originate from Ce 4$f$ electrons, while those located at around $-$20 eV below $E_{\rm F}$ are associated with Ce 5$p$ semicore electrons. It is noticeable that the LDOS shape of Ce atoms is very similar to that of H atoms in the energy range between $-$15 eV and $E_{\rm F}$, indicating a strong hybridization between Ce and H electronic states. In order to resolve the orbital characters of electronic states, we plot the partial density of states (PDOS) projected onto the Ce 5$p$-semicore and 4$f$/5$d$-valence states and the H 1$s$ state in Fig. 3(b). We find that the Ce 5$p$-semicore states are extended upward to reach up to $E_{\rm F}$, while the 4$f$- and 5$d$-valence states are distributed downward to about $-$10 and $-$13 eV below $E_{\rm F}$, respectively. Hence, these delocalized semicore and valence states hybridize well with the H 1$s$ state. Such a strong hybridization between Ce and H electrons is likely associated with the Ce-encapsulated spherical H-cage structure of CeH$_9$. Consequently, the electron charges of Ce and H atoms show covalent characteristics between the Ce and H$_1$/H$_2$/H$_3$ atoms [see the ``${\times}$" points in Fig. 2(a)]. Based on this covalent feature of the Ce$-$H bonds and the charge transfer from Ce to H atoms, we can say that the bonding nature between the Ce atoms to H$_{29}$ cages is characterized as a mixture of ionic and covalent.

As mentioned above, the synthesis of CeH$_9$ requires much lower pressures 80$-$100 GPa~\cite{ExpCeH9-Nat.Commun2019T.Cui,ExpCeH9-Nat.Commun2019-J.F.Lin} compared to that (${\sim}$170 GPa) for the synthesis of LaH$_{10}$~\cite{ExpLaH10-PRL2019,ExpLaH10-Nature2019}. Since La and Ce atoms belong to lanthanides or f-block elements with electron configurations [Xe]4$f^0$5$d^1$6$s^2$ and [Xe]4$f^1$5$d^1$6$s^2$, respectively, one expects that the lower synthesis pressure of CeH$_9$ would be caused by the presence of the delocalized Ce 4$f$ states [see Fig. 3(b)]. To confirm how the delocalized nature of Ce 4$f$ electrons contributes to the chemical precompression of H$_{29}$ cages, we optimize the structure of CeH$_9$ with varying pressure using the $f$-core scheme, where Ce 4$f$ electrons are considered as core electrons. The band structure and PDOS of CeH$_9$ calculated using the $f$-core scheme are displayed in Fig. S1. We find that the band dispersions of the Ce 5$d$ and H 1$s$ states change largely around $E_{\rm F}$ because their hybridizations with the Ce 4$f$ states are avoided in the $f$-core scheme. In Fig. 4(a), the lattice parameters computed using the $f$-core scheme are compared with those of the $f$-valence scheme as well as the experimental data~\cite{ExpCeH9-Nat.Commun2019T.Cui}. We find that in the pressure range between 100 and 160 GPa, both the $f$-core and $f$-valence schemes predict similar values for $a$ and $b$, close to the experimental values~\cite{ExpCeH9-Nat.Commun2019T.Cui}. However, the $f$-core scheme predicts larger $c$ values than the $f$-valence scheme and experiment~\cite{ExpCeH9-Nat.Commun2019T.Cui} by ${\sim}$6\% in the same pressure range. As a result, in contrast to both the $f$-valence scheme and experiment~\cite{ExpCeH9-Nat.Commun2019T.Cui}, the $f$-core scheme gives relatively larger values of the $c$/$a$ ratio between 100 and 160 GPa [see Fig. 4(b)]. These results indicate that the delocalized nature of Ce 4$f$ electrons plays an important role in determining the chemical precompression along the $c$ axis, while it hardly affects the chemical precompression in the $a-b$ plane.

In order to check whether the localized/delocalized nature of Ce 4$f$ electrons influences the dynamical stability of CeH$_9$, we calculate the phonon spectrum at 100 GPa using both the $f$-core and $f$-valence schemes. The calculated phonon spectrum of the $f$-core scheme exhibits negative frequencies in the whole Brillouin zone [see Fig. S2(a) in the Supplementary information], indicating that CeH$_9$ is dynamically unstable. On the other hand, the $f$-valence scheme shows that CeH$_9$ is dynamically stable without any negative-frequency phonon mode [see Fig. S2(b)]. Therefore, we can say that the delocalized nature of Ce 4$f$ electrons is necessary for stabilizing the clathrate structure of CeH$_9$.

\section*{Conclusion}
Our first-principles DFT + U calculations for CeH$_{9}$ have shown that (i) the Ce 5$p$ semicore and 4$f$/5$d$ valence states strongly hybridize with the H 1$s$ state, (ii) the charge transfer occurs mostly from Ce to H$_1$ and H$_2$ atoms, and (iii) the delocalized nature of Ce 4$f$ electrons is an essential ingredient in the chemical precompression of clathrate H$_{29}$ cages. The present results not only suggest that the bonding nature between the Ce atoms and H cages is characterized as a mixture of ionic and covalent, but also provide an explanation for the enhanced chemical precompression in CeH$_9$. It is thus proposed that the large chemical precompression of H-rich clathrate structures can be attained in rare-earth hydrides with delocalized 4$f$ electrons.

\section*{Methods}
Our DFT calculations were performed using the Vienna {\it ab initio} simulation package with the projector-augmented wave method~\cite{vasp1,vasp2,paw}. For the exchange-correlation energy, we employed the generalized-gradient approximation functional of Perdew-Burke-Ernzerhof (PBE)~\cite{pbe}. The 5$s^2$5$p^6$ semicore electrons of Ce atom were included in the electronic-structure calculations. For Ce 4$f$ electrons, we considered the effective on-site Coulomb interaction of U$_{\rm eff}$(=U$-J$) = 4 eV, where the Hubbard parameter U is 4.5 eV and the exchange interaction parameter $J$ is 0.5 eV~\cite{ExpCeH9-Nat.Commun2019T.Cui}. A plane-wave basis was used with a kinetic energy cutoff of 1000 eV. The ${\bf k}$-space integration was done with 12${\times}$12${\times}$8 $k$ points for the structure optimization and 24${\times}$24${\times}$16 $k$ points for the DOS calculation. All atoms were allowed to relax along the calculated forces until all the residual force components were less than 0.005 eV/{\AA}. Phonon calculations were preformed by a finite displacement method with the PHONOPY code~\cite{Phonopy} 

\phantomsection

\section*{Acknowledgements}
This work was supported by the National Research Foundation of Korea (NRF) grant funded by the Korean Government (Grants No. 2019R1A2C1002975, No. 2016K1A4A3914691, and No. 2015M3D1A1070609). The calculations were performed by the KISTI Supercomputing Center through the Strategic Support Program (Program No. KSC-2019-CRE-0183) for the supercomputing application research.

\section*{Author Contributions}
H.J., C.W., and S. Y. contributed equally to this work. J.H.C. designed the research; H.J. and C.W. performed the theoretical
calculations; all the authors analyzed the data and wrote the paper.\\

\section*{Additional Information}
\textbf{Competing financial interests} The authors declare no competing financial interests.

\clearpage
\onecolumngrid
\titleformat*{\section}{\LARGE\bfseries}

\renewcommand{\thefigure}{S\arabic{figure}}
\setcounter{figure}{0}

\vspace{1.2cm}

\section{Supporting Information for "Origin of enhanced chemical precompression in cerium hydride CeH$_{9}$"}
\begin{flushleft}
{\bf 1. Band structure and PDOS of CeH$_9$ in the $f$-core scheme}
\begin{figure}[ht]
\includegraphics[width=8cm]{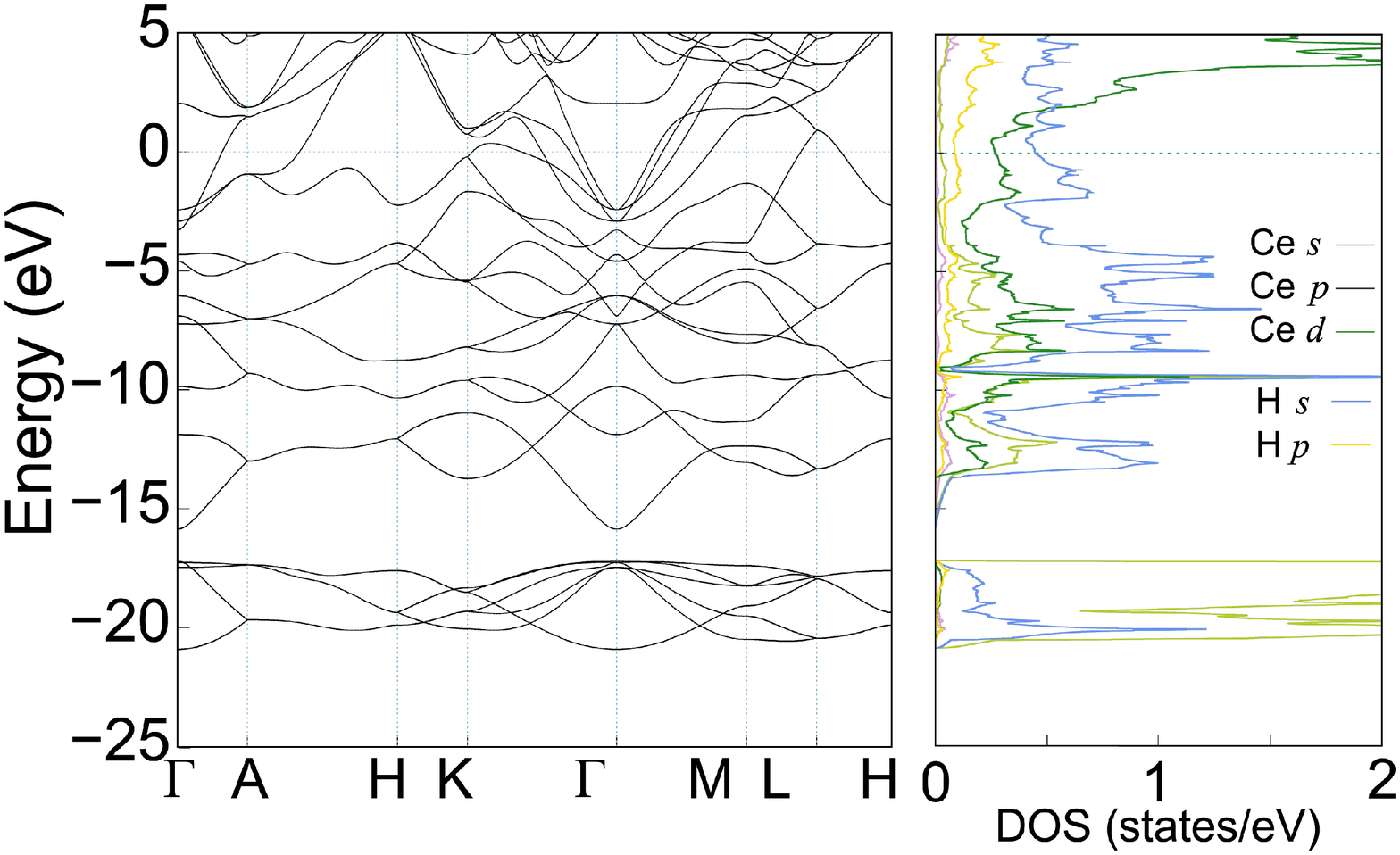}
\caption{ (a) Band structure and PDOS of CeH$_9$ at 100 GPa, calculated using the $f$-core scheme. Here, we use a special pseudopotential of Ce, in which f electrons are kept frozen in the core.}
\end{figure}

\vspace{1.2cm}

{\bf 2. Phonon spectra and DOS of CeH$_9$ in the $f$-core and $f$-valence schemes}
\begin{figure}[ht]
\includegraphics[width=\linewidth]{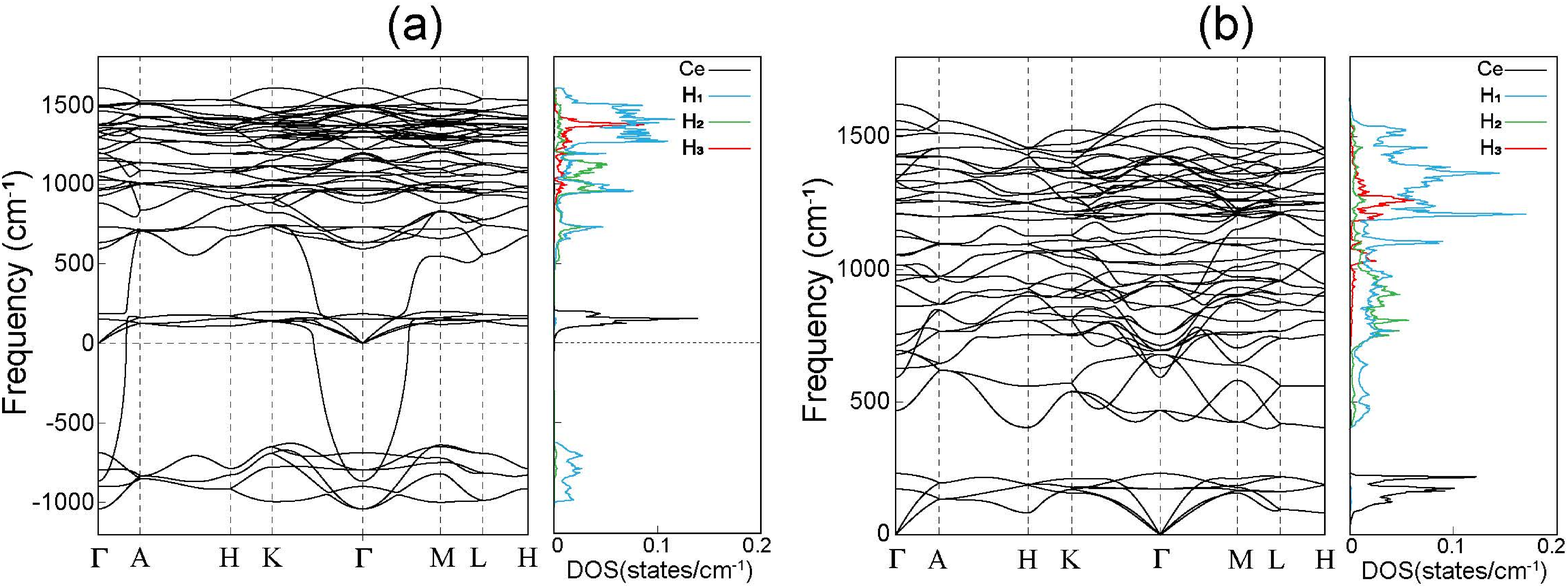}
\caption{ Phonon spectrum and DOS of CeH$_9$ at 100 GPa, calculated using the (a) $f$-core (b) $f$-valence schemes. The phonon DOS is projected onto Ce, H$_1$, H$_2$, and H$_3$ atoms. Here, the phonon spectra and DOS are obtained using the finite displace-ment method implemented in PHONOPY package, where the forces are calculated using the VASP code with a 2${\times}$2${\times}$2 supercell. }
\end{figure}
\end{flushleft}

\end{document}